\documentclass[10pt,twocolumn]{IEEEtran}
\usepackage{graphicx}           % Important!!
\usepackage{amsmath,amsfonts}
\usepackage{times,epsfig}
\usepackage{psfrag}
\usepackage{subfigure}
\usepackage{stfloats}
\usepackage{amssymb}
\usepackage{multirow}
\usepackage[justification=centering]{caption}
\usepackage{cite}
\usepackage{CJK}
\usepackage{url}
\usepackage{caption}
\usepackage{mathrsfs}
\usepackage{bm}
\usepackage{enumerate}
\usepackage{color,xcolor}
\usepackage{colortbl}
\usepackage{multirow}
\usepackage{array}
\usepackage{algorithmic}
\usepackage[lined,ruled,linesnumbered]{algorithm2e}
\usepackage{booktabs,tabularx}

\begin{document}
\pagenumbering{arabic}
\bibliographystyle{ieeetr}

\title{Federated Learning-Empowered AI-Generated Content in Wireless Networks}
 
\author{ Xumin Huang,  Peichun Li, Hongyang Du, Jiawen Kang, \emph{Member, IEEE}\\Dusit Niyato, \emph{Fellow, IEEE}, Dong In Kim, \emph{Fellow, IEEE}, and Yuan Wu, \emph{Senior Member, IEEE}
\IEEEcompsocitemizethanks{
\IEEEcompsocthanksitem X. Huang and P. Li are with State Key Laboratory of Internet of Things for Smart City, University of Macau, Taipa, Macau, China (e-mail: huangxu\_min@163.com; peichunli@um.edu.mo).
%%is with School of Automation, Guangdong University of Technology, Guangzhou 510006, China, and also with State Key Laboratory of Internet of Things for Smart City, University of Macau, Taipa, Macau, China (e-mail: huangxu\_min@163.com).
%%     Jiawen Kang, Weifeng Zhong and Shengli Xie are with School of Automation, Guangdong University of Technology, Guangzhou 510006, China.
H. Du, and D. Niyato are with the School of Computer Science and Engineering, Nanyang Technological University, Singapore (e-mail: hongyang001@e.ntu.edu.sg; dniyato@ntu.edu.sg). J. Kang is with School of Automation, Guangdong University of Technology, Guangzhou 510006, China (e-mail: kavinkang@gdut.edu.cn).  D. I. Kim is with the Department of Electrical and Computer Engineering, Sungkyunkwan University, Suwon 16419, South Korea (e-mail: dikim@skku.ac.kr). Y. Wu is with State Key Laboratory of Internet of Things for Smart City, University of Macau, Taipa, Macau, China, and also with Department of Computer and Information Science, University of Macau, Taipa, Macau, China (e-mail: yuanwu@um.edu.mo). %Dong In Kim is with the Department of Electrical and Computer Engineering, Sungkyunkwan University, Suwon 16419, South Korea (e-mail: dikim@skku.ac.kr).
}
}

\maketitle
\thispagestyle{empty}

\begin{abstract}
Artificial intelligence generated content (AIGC) has emerged as a promising technology to improve the efficiency, quality, diversity and flexibility of the content creation process by adopting a variety of generative AI models. Deploying AIGC services in wireless networks has been expected to enhance the user experience. However, the existing AIGC service provision suffers from several limitations, e.g., the centralized training in the pre-training, fine-tuning and inference processes, especially their implementations in wireless networks with privacy preservation.  Federated learning (FL), as a collaborative learning framework where the model training is distributed to cooperative data owners without the need for data sharing, can be leveraged to simultaneously improve learning efficiency and achieve privacy protection for AIGC. To this end, we present FL-based techniques for empowering AIGC, and aim to enable users to generate diverse, personalized, and high-quality content. Furthermore, we conduct a case study of FL-aided AIGC fine-tuning by using the state-of-the-art AIGC model, i.e., stable diffusion model. Numerical results show that our scheme achieves advantages in effectively reducing  the communication cost and training latency and privacy protection.  Finally, we highlight several major research directions and open issues for the convergence of FL and AIGC. 
\end{abstract}


%%%%%%%%%%%%%%%%%%%%%%%%%%%%%%%%%%%%
\section{Introduction}
%In this section, we introduce the concept of AI-generated content (AIGC) in wireless edg networks and the motivation behind utilizing federated learning-aided approaches to enhance its potential.

To generate a vast amount of high-quality digital content for Web 3.0 applications such as Metaverse, artificial intelligence generated content (AIGC) has emerged as a promising technology to adopt a variety of generative AI models for producing, handling and  modifying diverse data, e.g., text, image and audio. Due to the outstanding capability of content generation, AIGC achieves great potential in changing the lifestyle of humans and making tremendous progress in different domains. Furthermore, deploying AIGC services in wireless networks has been envisioned to enable users to have a real-time interactive environment,  context awareness support, personalized and engaging experience \cite{du2023Enabling}.  However, AIGC service provision  is still facing several challenging issues. On one hand, both model size and dataset size of a large-scale AIGC models are  extremely  large and cause critical difficulties to pre-train, fine-tune and infer the AIGC model among resource-constrained devices. For example, Chat Generative Pre-trained Transformer (ChatGPT) is one of the latest large language models trained on a huge amount of text data, i.e., 300 billion words (570 GB) and with over 175 billion parameters.  On the other hand,  data privacy has become a growing concern for users in the open wireless distributed computing  environment when they join AIGC services by using their local data. %In this regard, data protection laws such as the EU General Data Protection Regulation (GDPR) have presented stringent requirements for how to collect, process and store personal data in the context of AI applications.

Despite its great benefits and advantages, there have been several vital challenges in the AIGC processes, including  pre-training, fine-tuning, and inference as follows. 
\begin{itemize}
\item Since AIGC pre-training necessitates enormous computing power and  time, most of the devices with the limited resource capacity and power supply are difficult to join this process. Their data cannot be fully utilized.
\item AIGC fine-tuning transfers a pre-trained model to a wide range of downstream tasks. Fine-tuning aims to apply the pre-trained model's knowledge to new problems, i.e., downstream tasks. But the traditional fine-tuning process relies on continuous data collection from the users, which violates individual data privacy of users. 
\item Optimizations for AIGC inference suffer from imperfect information and uncertainties of network environment, where deep reinforcement learning (DRL) can provide a promising solution. For example, DRL was adopted in \cite{du2023Enabling} to orchestrate the multi-user AIGC inference. We further consider a cooperative multi-agent learning environment without centralizing the training data of all agents.
\end{itemize}

Facing the above challenges, federated learning (FL), as a most prevalent distributed training and collaborative learning framework, provides an effective approach to empower AIGC. FL enables distributed users/clients to collaboratively train a shared model while keeping all training data on the local storage.  In particular, FL provides a promising scheme for enabling the AIGC pre-training and fine-tuning  in a distributed manner. Compared with the centralized AIGC via cloud,   distributive AIGC via FL promotes collaboration among multiple clients and has advantages in  improving resource utilization efficiency, collecting fresh data for model training, reducing notable AIGC service delay and mitigating certain security and privacy threats. Thanks to FL, both efficiency and privacy of AIGC can be effectively improved. More specifically,
\begin{itemize}
	\item FL can be leveraged for AIGC pre-training by delegating a proper training task to a number of clients. Clients can join the pre-training process and contribute their data. 
	\item FL can facilitate AIGC fine-tuning  such that the clients can distributively fine-tune a pre-trained AIGC model in a privacy-preserving manner. %The existing parameter-efficient fine-tuning methods can be directly integrated into federated fine-tuning to reduce the communication overhead for the clients.
	\item Federated reinforcement learning enables each agent to only share the learning experience with other agents and accelerate the training process,   which is helpful to save the time and energy consumption for solving the multi-user AIGC inference problem.  
\end{itemize}
 
%It would be good to contrast the distributive AIGC via FL over the centralized AIGC via cloud, in view of the communication efficiency, resource utilization, age of information, latency, privacy and security issues, etc. 
 
% More data can be utilized for AIGC pre-training  and more clients are willing to participate in the fine-tune stage with strong privacy protection.  AIGC models are pre-trained well with the increasing data diversity and are widely fine-tuned for different downstream tasks. The multi-user service delivery problem of AIGC inference is also quickly tackled to accommodate diverse AIGC services while providing efficient quality-of-service management. This enables more users to generate more contents. 

However, the integration of FL and AIGC is not straightforward and poses several challenging issues. Accounting for the huge scale of an AIGC model with billions of parameters, it is prohibitive for a conventional client  to perform the pre-training of an entire AIGC mode or fine-tuning all model parameters. In addition, multiple iterations of local training and global aggregation in FL cause high resource consumption to AIGC services, which further aggravates challenges of deploying AIGC services in wireless networks.  In federated pre-training, each client is suitable to train a small part of the AIGC model. In federated fine-tuning,  we should limit the number of parameters which are fine-tuned by each client such that we can reduce the data communication of interacting with the parameter server. %With the emergence of federated reinforcement learning,  a critical problem about how to achieve the knowledge transfer across the agents without exchanging their private information raises. 
In federated fine-tuning and federated reinforcement learning for inference, proper  FL  designs can be discussed to accelerate the training convergence. The challenges motivate us to investigate how to exploit FL and its variants for empowering AIGC in different processes.

In this article, we focus on leveraging  FL for AIGC, and study how to decentralize the model training procedures regarding AIGC pre-training, fine-tuning and inference. We discuss the challenges of the existing AIGC, and propose instrumental  FL-based techniques to empower AIGC and implement the FL procedure in different approaches, i.e.,  parallel, split and sequential, according to different application conditions.  We compare the techniques and reveal the remarkable benefits of FL to AIGC.  
%Thanks to the proposed techniques, the amounts of available resource and data for pre-training a large AIGC model increase and pre-training on more data is useful for the AIGC model to enhance the diversity of generated content.  By combining federated learning with the existing parameter-efficient fine-tuning methods, we fine-tune the AIGC model among the clients in a privacy-preserving and communication-efficient manner.  AIGC fine-tuning becomes more cost-efficient to fit different downstream tasks while satisfying diverse user contexts. This enables users to generate personalized and customized content. In AIGC inference, a multi-user service delivery problem is  quickly tackled to accommodate diverse AIGC services such that it is  much more convenient for the users to generate the high-quality content on demand. 
Besides, we present a case study of FL-aided AIGC fine-tuning, where we consider the stable diffusion model, one of the state-of-the-art AIGC models, and study how to fine-tune the stable diffusion model on a neural style transfer task under the FL setting. Finally, we provide extensive discussions about opportunities and challenges of FL-empowered AIGC.

The main contributions of this article can be summarized as follows.
\begin{itemize}
	\item We propose a FL-empowered approach to introduce the FL-based techniques for the AIGC processes. To the best of our knowledge, this is the first work that considers the integration of FL and AIGC in wireless environment.
	%Based on the proposed federated learning-based techniques, we highlight the benefits of integrating federated learning with AIGC.
	\item We present a case study to show the potential of FL for fine-tuning the stable diffusion model. This case study can be straightforwardly extended to other AIGC fine-tuning scenarios. Compared with a traditional scheme, our scheme reduces both the communication cost and training latency while achieving the identical convergence performance to it. 
	\item  We outline research challenges and present potential solutions to the FL-empowered AIGC.  We discuss open directions on how economic theories and emerging technologies such as  semantic communication, blockchain and edge intelligence can be leveraged for the convergence of FL and AIGC.  
\end{itemize} 

%\section{Related Work}

\section{Deep Generative Models and AIGC Processes}
\begin{figure*}
	\centering
	\includegraphics[width=0.8\textwidth]{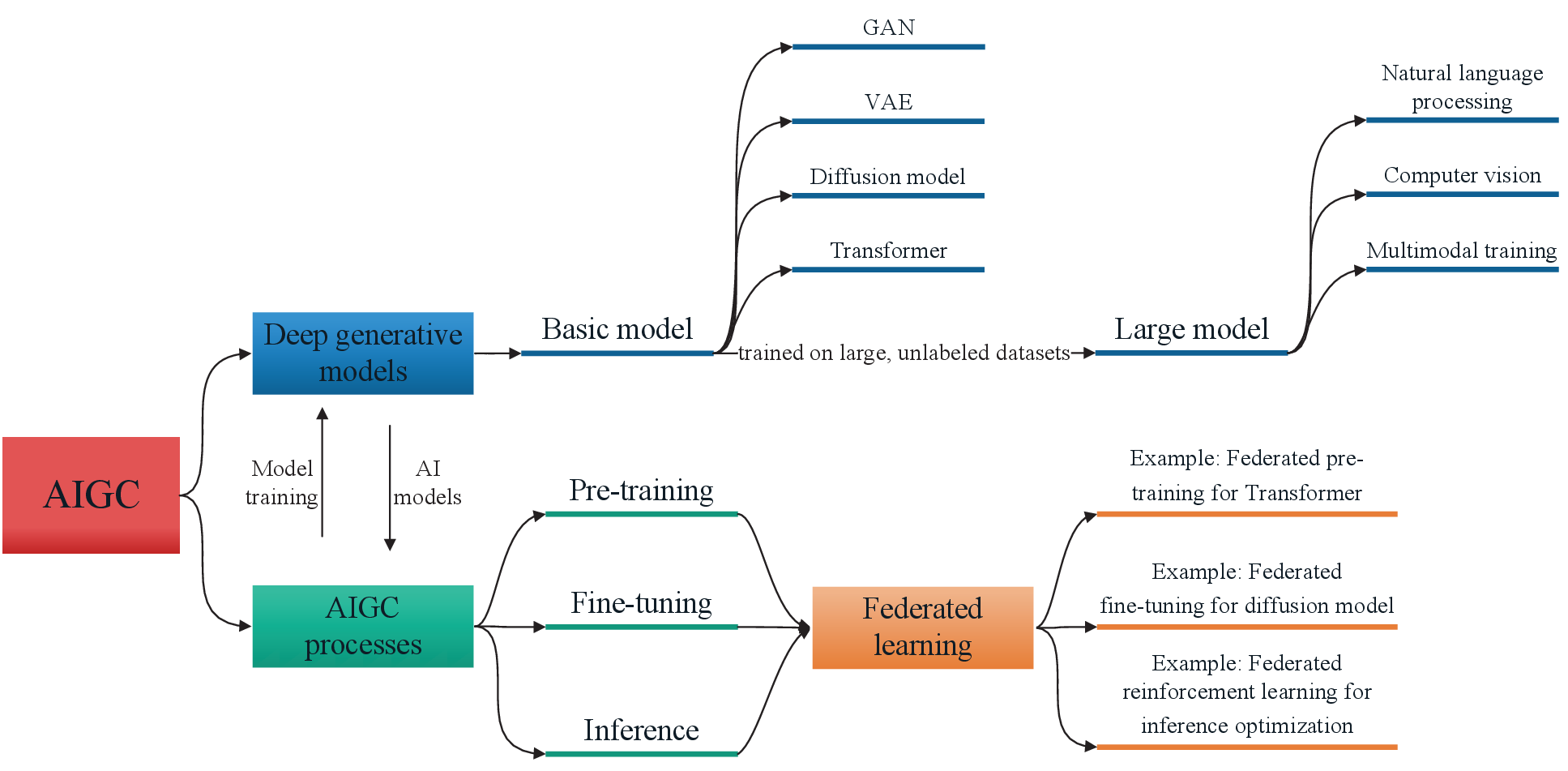}
	\caption{An overview of federated learning-empowered AIGC}
	\label{s0}
\end{figure*}

In this section, we present a short overview of FL-empowered AIGC, as shown in Fig.~\ref{s0}. We introduce the deep generative models and AIGC processes. The up-to-date research works of applying FL for the AIGC processes are also discussed.
\subsection{Fundamentals of Deep Generative Models}
As a key enabling technology for AIGC, deep generative model learns to generate predictions in the identical modality as the input data with one modality. The rapid development of deep generative models has witnessed several promising techniques and we pay attention to  Generative Adversarial Network (GAN), Variational Autoencode (VAE), diffusion model and Transformer. % We provide a brief introduction of the deep generative models as follows. 
We summarize technique details of the deep generative models and their comparisons in Table I.
\begin{itemize}
	
\item \emph{GAN}: As a unsupervised machine learning algorithm, GAN  aims to generate new data from the same space of the original data \cite{goodfellow2020}. GAN consists of two neural networks named by generator and discriminator, which are trained together in an adversarial manner. However, GAN cannot support multi-modal generation.  

\item \emph{VAE}: As a probabilistic generative model, VAE aims to maximize the probability of the generated output with respect to the input data \cite{kingma2019}. VAE is an adaptation from a standard autoencoder, transforms the data from a higher-dimensional to a lower-dimensional space and further tackles the critical problem of non-regularized latent space in the traditional autoencoder. Both GAN and VAE are two popular approaches that have been widely commonly used for generative modeling. Compared with GAN, VAE learns from labeled data and is easier to train.  

\item \emph{Diffusion model}:  Also known as diffusion probabilistic model, diffusion model is inspired by non-equilibrium thermodynamics proposed by Sohl-Dickstein et al. \cite{sohl2015}.  %Diffusion model  consists of a forward process that adds noise to the input and a reverse process that removes noise from the latent data. 
Diffusion model destructs training data by gradually adding Gaussian noise in the forward process, then reverses this process to generate the desired data from the noise.  Compared with VAE and GAN, diffusion model can achieve a better  learning performance but requires more training data, computational cost and time.  

\item \emph{Transformer}: Following an encoder-decoder structure, Transformer relies on the adaption of self-attention layers in  both the encoder and decoder \cite{vaswani2017}. % to perform the sequential data analysis
By capturing the relations among elements of the sequential data,  Transformer is suited to process sequence-to-sequence prediction tasks, and enables parallel training for both the data and model. Compared with the above generative models, Transformer achieves an advantage in performing the large-scale training. 
\end{itemize}

\begin{table*}[t]
	\renewcommand{\arraystretch}{2}
	\caption{Deep generative models for federated learning-aided AIGC}\centering \tabcolsep=5pt
	\begin{tabular}{p{0.15\columnwidth}p{0.25\columnwidth}p{0.45\columnwidth}p{0.45\columnwidth}p{0.45\columnwidth}}
		\hline
		\textbf{Model} & \textbf{Component} & \textbf{Feature} & \textbf{Comparison} & \textbf{Application}\\
		\hline
		 GAN  \cite{goodfellow2020} & Generator and discriminator & Two models are trained together in an adversarial manner in terms of a zero-sum non-cooperative game. & GAN adopts unsupervised learning and is sensitive to hyperparameter settings. & Image generation, restoration and  manipulation, super resolution and 3D shape reconstruction.	\\
		\hline
		VAE  \cite{kingma2019} & Encoder-decoder model  &  VAE tries to create  latent representations of training data in a probabilistic manner. & VAE utilizes  supervised learning and is easier to train and generate more coherent data. & Image denoising, image generation and audio synthesis.\\
		\hline
	Diffusion model \cite{sohl2015} & Likelihood-based model  & Add random noise to input data in the forward process and recover the data by removing the noise in the reverse process.
	%Define two Markov processes, i.e., a forward diffusion process and a reverse generative process. 
	& Diffusion model has the better learning performance but requires more training data, computational cost and time. & Computer version, natural language generation, and multi-modal generation. \\
		\hline
	 Transformer \cite{vaswani2017} &Sequence-to-sequence encoder-decoder model&  There are self-attention layers in both the encoder and decoder and each layer performs a multi-head attention mechanism. & Transformer enables multiple attention heads to work in parallel, and achieves an advantage in the scalability to large-scale datasets.&  Natural language processing and computer vision.\\	
		\hline
	\end{tabular}
\end{table*}

To enable the convenient use of AIGC,  pre-trained large models, i.e., foundation models, are trained on large, unlabeled datasets and developed to adapt to a wide range of downstream tasks.  In recent years, leading AI industries  have presented a variety of pre-trained large models for AIGC, including the models for natural language processing such as BERT by Google and GPT by OpenAI, the models for computer vision such as Florence by Microsoft, the models for multimodal training such as Dall-E 2 by OpenAI,  Imagen by Google, and stable diffusion by Stability AI. We provide more details of GPT and stable diffusion as follows. 

Many foundation models in the natural language processing domain rely on the Transformer-based architecture for training, due to the remarkable learning ability and parallelism of Transformer. 
In this regard, GPT refers to a series of generative pre-Trained Transformer models released by OpenAI and has evolved from the first version GPT-1 in 2018 to the latest model GPT-4 in 2023. GPT 3.5 is an intermediate version between GPT-3 and GPT-4.   As a popular AI chatbot based on GPT-3.5, ChatGPT is developed as a variant of GPT with chatbot functionality, and is fine-tuned by jointly utilizing supervised learning and  reinforcement learning to improve the conversational abilities of the chatbot\footnote[1]{https://openai.com/blog/chatgpt}.  Due to the fast-rising popularity, ChatGPT reportedly has over 100 million active users globally. In addition, diffusion model provides a cutting-edge approach for text-conditioned image generation and  stable diffusion as one of the well-known foundation models  have attracted extensive attention. Stable diffusion is an  open-source visual generative foundation model released by Stability AI in 2022, and has been commonly used for generating high-quality images in massive applications. Stable diffusion is one of the most flexible AI image generators to create AI-generated images and modify the images on demand, e.g., inpainting and super-resolution, according to the text prompts\footnote[2]{https://stability.ai/blog/stable-diffusion-v2-release}. Besides, running a stable diffusion model also requires relatively lower memory, e.g., 16 GB of DDR4 or DDR5 RAM, and GPU usage, e.g, 10 GB VRAM.

%GPT-3.5 and GPT-4
%
%
%ChatGPT is a member of the generative pre-trained transformer (GPT) class of language models. It is a task-specific GPT that was fine-tuned to target conversational usage, and was originally built upon an improved version of OpenAI's GPT-3 model known as "GPT-3.5".[7]
%
%The fine-tuning process leveraged both supervised learning as well as reinforcement learning
%
%
%
%ChatGPT is a generative pre-trained transformer model that is fine-tuned from a model in the GPT-3.5 series
%
%is a variant of GPT
%
%
%The foremost architectural distinction is that in a transformer’s encoder-decoder model, BERT is the encoder part, while GPT-3 is the decoder part.
%
%Chat
%
%Transformers revolutionized AI generation
%
%because of their learning ability and parallelism.
%
%are the foundation for many state-of-the-art NLP systems.

%\begin{table*}[t]
%	\renewcommand{\arraystretch}{2}
%	\caption{Deep generative models for AIGC}\centering \tabcolsep=5pt
%	\begin{tabular}{|p{0.15\columnwidth}|p{0.1\columnwidth}|p{0.7\columnwidth}|p{0.7\columnwidth}|}
%		\hline
%		\textbf{Model} & \textbf{Year} & \textbf{Basic component} &  & \textbf{Feature}\\
%		\hline
%		 GAN & 2014 & Generator and Discriminator & \\
%		\hline
%		VAE & 2014 & Encoder-Decoder model  &  \\
%		\hline
%	Diffusion model& 2015& Likelihood-based model  & multi-modal generation \\
%		\hline
%	 Transformer & 2017 & &\\
%		\hline
%		 CLIP & 2021 & & \\
%		\hline
%	\end{tabular}
%\end{table*}

\subsection{AIGC Processes}
In the following, we discuss the challenging issues and current solutions of AIGC processes, including pre-training, fine-tuning and inference.

\subsubsection{Pre-training}
AIGC  depends on  large-scale pre-trained models with several billions of parameters. Existing pre-training methods require an available large-scale dataset and a huge investment in hardware (e.g., thousands of GPUs) and time (e.g., several days). For example, training a GPT-3 model with over 175 billion parameters requires more than 4 months with 1,000 GPUs and the training cost is estimated from 4.6 to 12 million US dollars for a single training run. The big companies like Google and OpenAI can afford but individuals and small companies are difficult to help pre-train a large model although they have available training data.  Effective collaboration among different parties should be explored to facilitate the pre-training process and improve the diversity of pre-training data.

To improve the pre-training performance, effective schemes are provided to alleviate the unbearablel computational time and expense. Take the Transformer-based foundation models as an example. Proper pre-training designs are introduced to  make well use of the inherent parallelism in Transformers. %which is helpful to  facilitate the pre-training process of Transformer-based foundation models. In addition, 
Model architecture change by applying the switchable transformer blocks to drop some Transformer layers for each mini-batch \cite{zhang2020} is proposed to train the models at a faster rate. 
%Due to the great advantages of enabling collaborative learning while preserving data privacy, 
FL has been leveraged to enhance the pre-training processes of BERT and ImageNet. The authors in \cite{tian2022} combine FL with split learning to prevent collecting the tremendous training data and permit resource-constrained clients to join the  BERT pre-training. Split learning enables each client to train a computation-lightweight part of the model as a client-side model and FL further enables all clients to collaboratively train the shared client-side model in a privacy-preserving manner.  Federated self-supervised learning is integrated with the existing masked image modeling methods to  facilitate the  ImageNet pre-training \cite{yan2023}.

%To overcome the label deficiency problem in medical imaging and simultaneously learn visual representations from decentralized unlabeled data, federated self-supervised learning is integrated with the existing masked image modeling methods to  improve the pre-training performance of ImageNet \cite{yan2023}. %The comprehensive scheme achieves an improved out-of-distribution generalization capability.

\begin{table*}[t] \centering
	\caption{Comparison of different fine-tuning methods for federated learning-aided AIGC}\label{table:compare} \centering
	\begin{tabularx}{\textwidth}{lXXXX}
		\toprule
		\multirow{2}{*}{Method} & \multirow{2}{*}{Description} & \multirow{2}{*}{Feature} & \multicolumn{2}{c}{Combining with federated learning} \\
		\cmidrule{4-5}
		&                            &                          & \multicolumn{1}{c}{Pros}              & \multicolumn{1}{c}{Cons}              \\
		\midrule
		Traditional
		& Fine-tune the full or partial weights of a pre-trained model. & Very large gradient size for model update (1$\sim$10GB)  &Simple and effective & Likely causing over-fitting when learning with few training samples\\
		\midrule
		LoRA \cite{hu2021} & Add extra bypass layers with few parameters for low-rank adaption. & Moderate gradient size ($\sim 100$ MB), moderate training data ($\sim$ 30 samples)& Exploiting decomposition matrices for \textit{communication and data efficient} tuning & Deteriorate the diversity of the generated content\\
		\midrule
		DreamBooth \cite{ruiz2022} & Create a personalized model for a specific subject under various contexts.& Large gradient size ($\sim 2$ GB), very little training data ($3\sim5$ samples) & Leverage the semantic prior  for \textit{data-efficient} training & Communication-intensive fine-tuning for each single subject  \\
		\midrule
		Textual inversion \cite{gal2022}
		& Capture novel concepts for personalized image generation via embedding tuning.  & Small embedding vector ($\sim 1$  KB), very little training data ($3\sim5$ samples) & Manipulate the word embedding for \textit{communication and data efficient} fine-tuning & Computation-intensive fine-tuning for each word embedding\\
		\midrule
		ControlNet \cite{zhang2023}
		& Introduce extra conditional input for controlling the generated content. &  Large gradient size ($\sim 1$ GB), extra inference cost& Improve the \textit{scalability} of the based model from diverse data over edge networks & Communication-intensive; incompatible with joint training of multiple controllers \\
		\bottomrule
	\end{tabularx}
\end{table*}
\subsubsection{Fine-tuning}
Before the practical use of a pre-trained AIGC model,  fine-tuning is performed to properly enhance the model's performance on a downstream task and adjust the model's parameters to suit the new data, e.g., in a new domain.

We specifically consider the diffusion-based text-to-image generative model and introduce  baseline methods for AIGC fine-tuning.  The traditional fine-tuning method is to directly fine-tune the full or partial weights of a pre-trained model. Many advanced fine-tuning methods have been presented to overcome the limitations of the traditional method. Low-rank adaptation (LoRA) fine-tuning is proposed by Microsoft to explore low-rank adaptation to facilitate the fine-tuning process in a storage and computation efficient manner \cite{hu2021}. %The method freezes all weights of a pre-trained model,  injects trainable rank decomposition matrices in each transformer block, and finally only optimizes the smaller rank decomposition matrices of the dense layers’ change during the fine-tuning. Without the calculation of the gradients of the most model weights, LoRA fine-tuning saves the GPU and storage usage while accelerating the fine-tuning process. 
In Section IV, we apply this method to  fine-tune a stable diffusion model on a customized dataset.  DreamBooth fine-tuning \cite{ruiz2022} is proposed by Google to bind a unique identifier with an interesting subject in a few (typically 3-5) input images, and embed the subject within the pre-trained diffusion model's output domain to generate new images of the subject. %Although the prior-preservation loss is introduced to generate diverse images,  But this method has a limited data diversity and requires to optimize numerous parameters. 
Textual inversion fine-tuning \cite{gal2022} is proposed by NVIDIA to find new pseudo-words that capture high-level semantic information and  fine-grained visual details in the textual embedding space of a frozen latent diffusion model. %Each word in an input string of the text encoder is converted to a token corresponding to an embedding vector and the embedding vectors are optimized  according to  a visual reconstruction objective. 
%Similarly, this fine-tuning method only utilizes 3-5 reference images. 
ControlNet \cite{zhang2023} adds an adapter model to a frozen pre-trained diffusion  model, and train the adapter model to let the diffusion model support additional input conditions. %To achieve image editing, the diffusion model is augmented with some control elements, e.g., edge map and  segmentation map. 
We have a brief introduction of the above fine-tuning methods and compare their gradient  and  training data sizes in the first three columns of Table II.

%However, fine-tuning the AIGC model requires a collection of potentially sensitive data from many users. This inevitably causes privacy threats to users and hinders their participation willingnesses.  Privacy-preserving and distributed fine-tuning is exactly necessitated to avoid collecting their private data while enabling collaborative learning among the users within the acceptable time. To tackle the problem, federated learning has begun to be applied for decentralizing the fine-tuning process of a foundation model. For example, federated prompt-tuning \cite{10095356}  takes both advantages of federated learning and prompt-tuning to simplify the fine-tuning stage of CLIP by letting all clients only learn trainable prompts  regarding a downstream task in a decentralized manner.

\subsubsection{Inference}
A fine-tuned AIGC model is deployed in wireless networks to provide AIGC inference services for users. %According to dynamic network states and diverse user preferences,  the decisions on  resource allocation, incentive design and user associations, have to be made.  When the services have insufficient prior knowledge, centralized decision making cannot be achieved and this leads to  a decision making problem under incomplete information. 
Several works have been presented to improve the inference efficiency and provide considerable payments for the inference services.  %The authors in \cite{du2023Distributed} introduce a distributed inference method for multiple users with the diffusion model. The proposed method is based on grouping the users according to their inference requirements and performing some shared denoising steps  for each user group. 
To incentivize AIGC service providers, a diffusion-based contract theory is employed to tackle the problem on how a user designs proper contracts for different AIGC service providers with different types \cite{liu2023}. Deep reinforcement learning is adopted  to tackle the inference service matching problem under incomplete information  \cite{du2023Enabling}. Reinforcement learning algorithms show great potential to overcome the incomplete information condition when the inference services have insufficient prior knowledge. Furthermore, we need to prevent each agent from directly sharing any private information in the cooperative multi-agent learning environment.

According to the overview of AIGC, we realize that FL has begun to be applied for decentralizing the model training procedures regrading the AIGC processes. However, the integration of FL and AIGC is not straightforward.  We take fine-tuning an AIGC model in a federated manner as an example. For the privacy protection, the federated fine-tuning methods are motivated  but technical challenges remain unresolved. For example, the combination of FL and LoRA is helpful to exploit the decomposition matrices for achieving the communication and data-efficient fine-tuning while this may deteriorate the diversity of the generated content. We summarize the  pros and cons of combining the fine-tuning methods with FL for AIGC in the last two columns of Table II. %We further discuss several challenges in the integration of federated learning and AIGC. We take fine-tuning an AIGC model in a federated manner as an example. 

\section{Federated Learning-Aided AIGC}
In this section, we introduce features, technique details and discussions of FL-aided AIGC.
\subsection{Features of Federated Learning Designs for AIGC}
We discuss the necessitated features of the FL designs for AIGC fine-tuning when both the FL and AIGC are deployed in wireless networks. 
\begin{itemize}
	\item \emph{Data-efficient}:  For local model training of each client,  the amount of downstream data is properly decided to overcome the overfitting problem and  restrict the number of local training iterations. 
	%while not sacrificing the generalization ability of a  pre-trained  model.
	\item \emph{Hardware-friendly}: Fine-tuning methods are selected and adjusted to reduce the GPU VRAM usage and make GPU-based computation more efficient. %by leveraging inter-GPU and intra-GPU parallelism, which is important to enable a hardware-friendly federated learning on the client side.  
	\item \emph{Parameter-efficient}: Parameter-efficient  fine-tuning techniques aim to train a small portion of the model parameters while keeping the rest frozen to cut down the computation workloads, time and communication cost. %This is beneficial to cut down the computation workloads, run time and communication cost for the resource-constrained clients with the limited battery lifetime and  wireless bandwidth.
	%which are difficult to submit massive local model updates with the limited battery lifetime and  wireless bandwidth.
\end{itemize}

%LoRA can save the weights for the new layers as a single file that weighs in at ~3 MB in size. This is about one thousand times smaller than the original size of the UNet model!

%\begin{figure*}[!htbp]
%	\setlength{\belowcaptionskip}{-0.3 cm}
%	\centering
%	\subfigure[Parking lot operator 1.]{
%		\label{s1a} %% label for first subfigure
%		\begin{minipage}[!htbp]{0.5\linewidth}
%			\centering
%			\includegraphics[width=1.0\textwidth]{FL-V4a}
%	\end{minipage}}
%	\subfigure[Parking lot operator 2.]{
%		\label{s1b} %% label for second subfigure
%		\begin{minipage}[!htbp]{0.5\linewidth}
%			\centering
%			\includegraphics[width=1.0\textwidth]{FL-V4b}
%	\end{minipage}}
%	\subfigure[Parking lot operator 3.]{
%		\begin{minipage}[!htbp]{0.5\linewidth}
%			\centering
%			\label{s1c} %% label for third subfigure
%			\includegraphics[width=1.0\textwidth]{FL-V4c}
%	\end{minipage}}
%	\caption{Comparison of training loss under different approaches.}
%	\label{s1}
%\end{figure*}

\begin{figure*}
	\centering
	\includegraphics[width=0.8\textwidth]{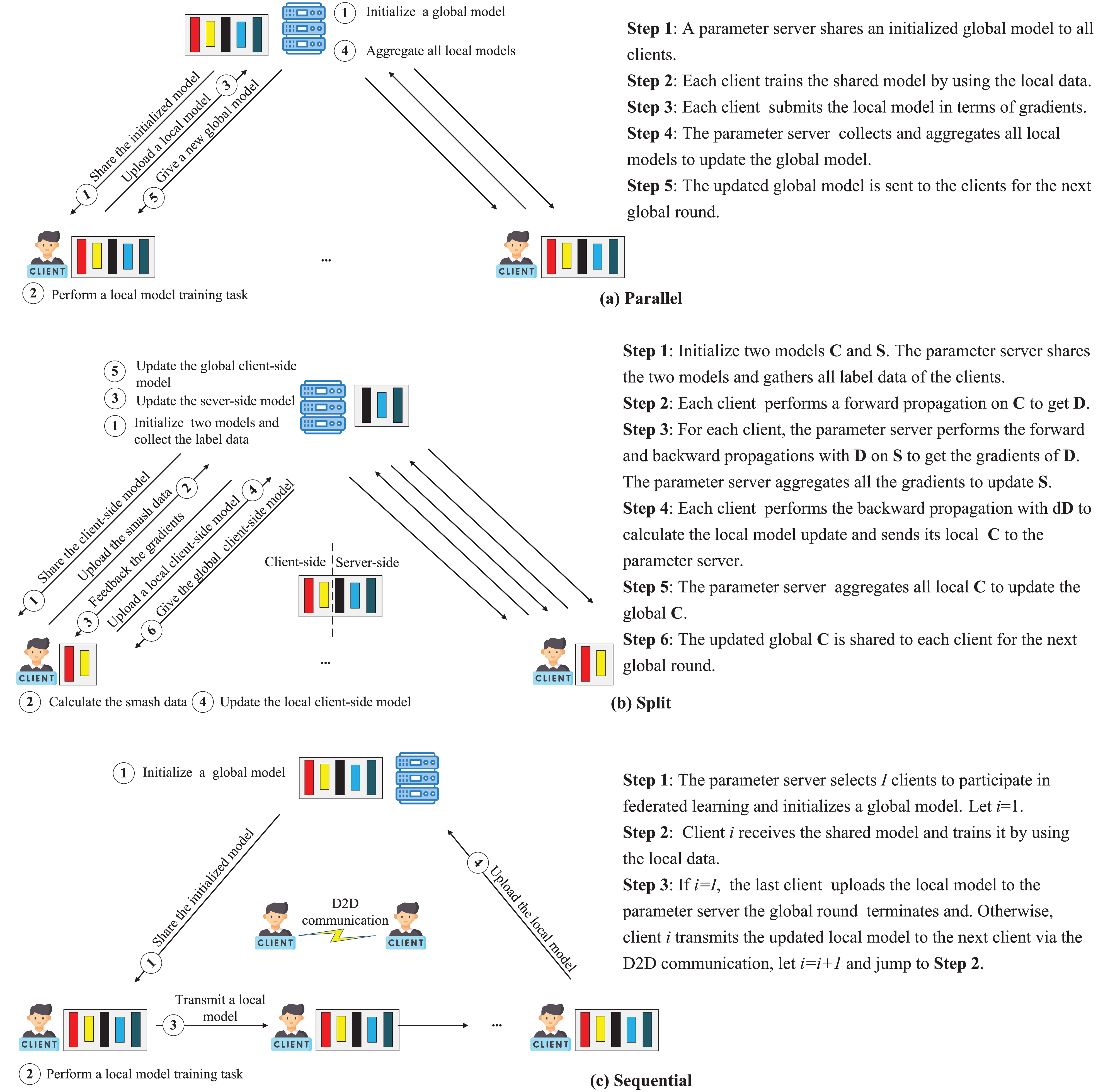}
	\caption{Federated learning-based techniques for AIGC}
	\label{s1}
\end{figure*}

%\begin{figure}[t!]
%	\setlength{\belowcaptionskip}{-0.3 cm}
%	\centering
%	\subfigure[Comparison of objective utility of the offloading user with respect to different $\alpha^j$ and schemes.]{
%		\label{s1a} %% label for first subfigure
%		\includegraphics[width=0.45\textwidth]{FL-V6a}}
%	\subfigure[Comparison of utility of the PVs with respect to different PV types and schemes.]{
%		\label{s1b} %% label for second subfigure
%		\includegraphics[width=0.45\textwidth]{FL-V6b}}
%	\caption{Performance comparison of utility formation under different incentive schemes.}
%	\label{s1}
%\end{figure}

\subsection{Federated Learning-Based Techniques for AIGC}
We propose FL and the variants to empower AIGC, as shown in Fig.~\ref{s1}. %Before participating in federated learning, each client collects sufficient data via the data collection techniques such as data trading and crowdsourcing, and utilizes the data for  local model training. 
In the conventional FL, all clients train an entire learning model in a parallel approach. This may hinder the participation willingness of lightweight clients that are difficult to train the computation-intensive model, and degrade the convergence rate  when a large number of clients with heterogeneous data sizes and computing capability participate in FL. Thus, we propose to implement FL in other approaches, i.e., split  and  sequential, according to the specific application scenarios. We aim to extend the adoption of FL in the AIGC processes, and make FL designs better suited for different application purposes. We provide more details as follows.
%In the following, we introduce the federated learning and the variants for AIGC.   

%\subsubsection{Federated learning in a parallel approach}: The implementation of the conventional federated learning is shown as follows. 
%\begin{itemize}
%	\item Step 1: global model initialization. There is a parameter server with a central position that transmits a shared model (i.e., global model) to all clients.
%	\item Step 2: local model training. Each client receives the global model and locally trains the model by using the local data.
%	\item Step 3: local model uploading. Each clients submits the local model in terms of gradients  to the parameter server. 
%	\item Step 4: local model aggregation. The parameter server collects and aggregates all local models to update the global model. 
%	\item Step 5: global model distribution. The updated global model is sent back to the clients for the next global round.
%\end{itemize}
%The above Steps 2-5 repeats until the global model converges.

\subsubsection{FL in a parallel approach}: The implementation of the conventional FL is shown in Fig. 2(a). In each global round (i.e., communication round), Steps 2-5 are performed and the steps repeat until the global model converges.

\subsubsection{FL in a split approach}: The implementation depends on the joint utilization of FL and split learning. The entire AIGC model is split into two sub-models: client-side model $\bf{C}$  and server-side model $\bf{S}$.  The split approach provides an opportunity for general clients to collaboratively train a large AIGC model since the client-side model is computation-lightweight and acceptable for the general clients with limited computing capability. In a global round of FL, each client receives the initialized weights of the client-side model and  trains it by using the local data. The parameter server has dual responsibilities: i) helps each client  update its client-side model and updates its server-side model by using split learning, ii) aggregates all client-side models by using the FedAvg algorithm, which assigns higher weights to the clients with more local data. Here, output of the last layer of a client-side model is named by smash data $\bf{D}$. We provide more details of the global round in Fig. 2(b).

\subsubsection{FL in a sequential approach}: The implementation depends on device-to-device (D2D) based model propagation \cite{hosseinalipour2022multi} among all clients. All local model training tasks are performed among the clients one by one. A client takes a  turn to play as a relay training node, which receives a local model from its previous client, train the local model by using the local data, and pass the updated local model to the next client. The model propagation between any two adjacent clients is quickly performed by using D2D communications to improve spectral and energy efficiency of cellular networks.  We provide more details of the global round in Fig.~2(c).  

%\subsubsection{Federated learning in a sequential approach}: The implementation depends on device-to-device (D2D) based model propagation \cite{hosseinalipour2022multi} among all clients. All local model training tasks are performed among the clients one by one. A client takes a  turn to play as a relay training node, which receives a local model from its previous client, train the local model by using the local data, and pass the updated local model to the next client. The model propagation between any two adjacent clients is quickly performed by using D2D communications to improve spectral and energy efficiency of cellular networks.  We provide more details of the global round $t$ as follows.  
%\begin{itemize}
%	\item Step 1: The parameter server schedules $I$ clients to participate in federated learning and initializes a shared model ${\bf}{W}_t$, which is first transmitted to client $i=1$.	
%	\item Step 2:  Client $i$ receives a local model ${\bf}{W}_{i,t}={\bf}{W}_{i-1,t}$. Particularly, we let ${\bf}{W}_{0,t}$ equalizes the initialized model. Based on the local data, the client performs the forward and backward propagations to calculate the gradients and further update ${\bf}{W}_{i,t}$.
%	\item Step 3: If $i>I$, the global round $t$ terminates and the last client $i=I$ uploads the local model to the parameter server for the next global round. Otherwise, client $i$ transmits the updated local model to the next client via the D2D communication, let $i=i+1$ and jump to Step 2. 
%\end{itemize}
\subsection{Discussions on Federated Learning-Empowered AIGC}

%we should have 1-2 paragraphs to discuss the lessons learned to compare these three approaches (parallel, split, and sequential). We can highlight the advantages and disadvantages, the situations that they can be applied, convergences, scalability, privacy preservation strength, suitability for which types of generative models, etc. 

FL can be performed in different approaches to empower AIGC. The parallel approach is easily applied and can resist a certain proportion of security threats such as model poisoning attacks through the gradient averaging operation. However, it presents a relatively higher computational requirement for the participating clients since each client needs to train the entire AIGC model. The convergence performance of the original FL cannot be guaranteed with the increasing number of the clients. The split approach prevents the clients from undertaking the heavy computation workloads, and enables the resource-constrained clients to join the FL procedure by letting them only train a computational-lightweight part of the model. The split approach is suitable for training the models with Transformer-based architectures since the models are conveniently split into two parts with different computation workloads. However, the split approach may be inapplicable for training some special AIGC models since the parameter server could simultaneously require the input data and label data of each client to perform the backward propagation for updating the client-side model. For example, FL in a split approach is not compatible with the training of a conventional denoising diffusion probabilistic model, which necessitates the original image data of the clients for each training iteration. At this time, sharing the local data of the clients with the parameter server will violate the original privacy-preserving rule of FL. The sequential approach is convinced to greatly improve the convergence rate of  FL. However, the sequential approach could aggravate some security threats to the FL procedure. Due to the model propagation among the clients, the sequential approach makes the FL procedure more susceptible to model poisoning attacks and gradient leakage attacks.

%Thanks to the proposed techniques, the amounts of available resource and data for pre-training a large AIGC model increase and pre-training on more data is useful for the AIGC model to enhance the diversity of generated content.  By combining federated learning with the existing parameter-efficient fine-tuning methods, we fine-tune the AIGC model among the clients in a privacy-preserving and communication-efficient manner.  AIGC fine-tuning becomes more cost-efficient to fit different downstream tasks while satisfying diverse user contexts. This enables users to generate personalized and customized content. In AIGC inference, a multi-user service delivery problem is  quickly tackled to accommodate diverse AIGC services such that it is  much more convenient for the users to generate the high-quality content on demand.

%The proposed federated learning-based techniques encourage collaboration among clients while preserving their data privacy in the model training procedures regarding the AIGC processes.  
The proposed FL-based techniques are helpful to facilitate the development and deployment of the AIGC models with more data utilization for pre-training, less communication cost for fine-tuning and higher-level optimizations for inference. We summarize the benefits of FL to AIGC as follows.
%We summarize the benefits of federated learning for AIGC as follows. 
\begin{itemize}
	\item \emph{Data diversity for AIGC pre-training}: The diversity of pre-training data increases
	after FL enables the clients to help pre-train a large AIGC model. %The diversity of pre-training data increases and this is significant for the creation of a generative pre-trained model, which further encourages diversity of the generated content.
	%The diversity of generated content also increase. potentially producing more diverse content.   
	\item \emph{Communication saving for AIGC fine-tuning}: FL and the parameter-efficient fine-tuning methods work together to improve the communication efficiency in the AIGC fine-tuning. %For example, we adapt the pre-trained AIGC model to a downstream tasks by only tuning and training a limited number of soft prompts with the small data sizes such that the fine-tuning stage is accelerated and remarkably saves the communication cost.  The AIGC model can be flexibly fine-tuned to process different downstream tasks and meet specific user needs. This enables users to customize AIGC services and generated personalized content on demand.
	\item  \emph{Knowledge transferring for AIGC inference}: Federated reinforcement learning encourages the agents to use the knowledge of other agents,  which is beneficial to accelerate the training process of the multi-agent learning.
\end{itemize}

\section{Case Study: Federated Fine-tuning for a Stable Diffusion Model}
\begin{figure*}
	\centering
	\includegraphics[width=0.8\textwidth]{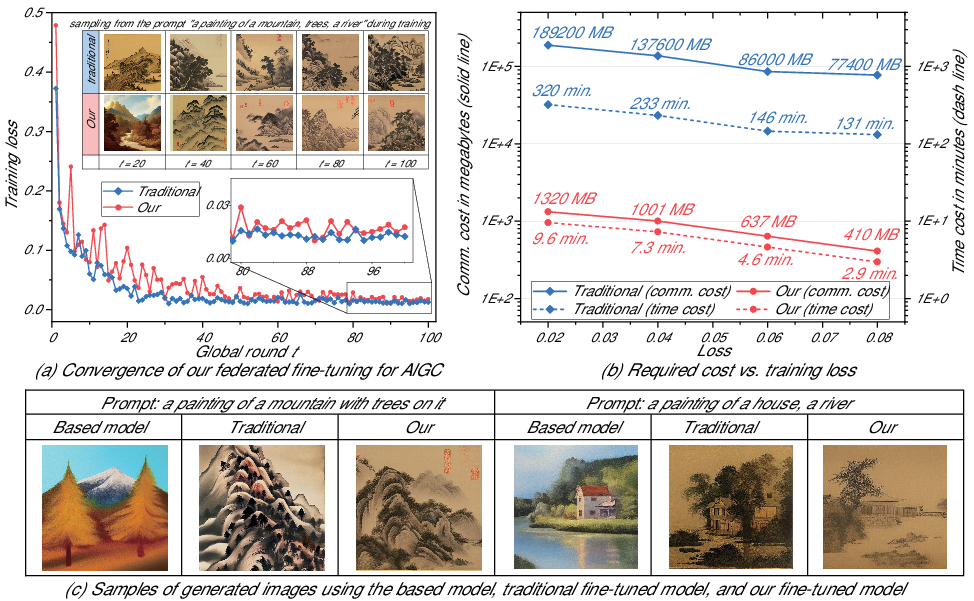}
	\caption{Federated fine-tuning for a stable diffusion model}
	\label{fig:case_study}
\end{figure*}

In this section, we present a detailed case study of federated fine-tuning for a stable diffusion model.  We perform the FL procedure in a sequential approach for fine-tuning the stable diffusion model in \cite{rombach2022high}. We aim to generate traditional Chinese ink paintings by leveraging the LoRA-based neural style transfer technique.  We design a federated fine-tuning framework where multiple clients collaboratively customize the general image synthesis model for creating a unique artistic style and D2D based model propagation is utilized to accelerate the training convergence for them.

We consider a training scenario with a set of clients and a parameter server.
%in the cellular network, where a base station equipped with an edge server is located in the center. 
Let $\bm{w}$ denote the weight of the foundation stable diffusion model, and $\bm{v}$ denote the weight of the additional bypass introduced by LoRA. During the fine-tuning process, we fix the base weight $\bm{w}$ and only train the additional weight $\bm{v}$. The training procedure of the federated fine-tuning with the D2D communications is described as follows.

\begin{itemize}
	\item Step 1: The parameter server initializes an index set ${\cal S}$ that equalizes the original set of the clients, and randomly selects a client $i \in {\cal S}$. In the global round $t$, the parameter server transmits the bypass weights $\bm{v}_{t}$ to client $i$.
	\item Step 2: The selected client $i$ performs a local model training task with local training samples and produces the new weights for the bypass model $\bm{v}_{t,i}$. After that, client $i$ is split from the index set, ${\mathcal S}={\mathcal S}-{i}$.
	\item Step 3: The current global round terminates if ${\cal S}=\varnothing$ and go to the next global round $t=t+1$. Otherwise, client $i$ selects a new client $i^\prime \in {\cal S}$ as the next relay training node. When the local weights $\bm{v}_{t,i}$ is transmitted from client $i$ to client $i^\prime$, we assign $i=i^\prime$ and jump to Step 2. 
	\item Step 4: The above steps 1-3 are repeated $T$ global rounds when $t>T$. Then the weights of the base model $\bm{w}$ and additional bypass $\bm{v}_T$ are merged together to form the weights of the fine-tuned personalized model $\bm{w}^{\prime}$, which is utilized to generate the customized content. Note that the inference cost of  $\bm{w}^{\prime}$ is identical to that of $\bm{w}$.
	\end{itemize}
%After $T$-round iterations of the above three steps, we obtain the final weight $\bm{v}_T$. Then, the weights of the base model $\bm{w}$ and additional bypass $\bm{v}_T$ are merged to form the fine-tuned personalized model $\bm{w}^{\prime}$, the inference cost of which is identical to that of $\bm{w}$. During deployment time, we employ the fine-tuned model $\bm{w}^{\prime}$ to generate the customized content.

To evaluate the performance of our proposed federated fine-tuning scheme, we collected 20 pieces of ancient Chinese ink paintings by different masters and cropped them to a resolution of $512 \times 512$ pixels. For simplicity, the total number of clients is 5 and each client has 4 training samples. For the hyper-parameter setting, we use a number of low-rank dimensions of 8 and let the total number of global rounds $T=100$. We compare the performance between our scheme and a traditional fine-tuning scheme, which trains all parameters of the U-Net in the stable diffusion model. To reduce the training cost, we apply the half-precision floating-point for the parameter representation. Specifically, the required data sizes for D2D parameter transmission of the proposed and traditional schemes are 9.1 and 1,720 megabytes, respectively.

Figure \ref{fig:case_study}(a) shows the training loss over the global round between these two schemes. We observe that the traditional scheme converges faster than the our scheme in the first 60 rounds and our scheme achieves comparable training loss to the traditional scheme after 100 global rounds. During the training, we periodically sample and record the output image of the generator every 20 rounds with the same prompt. The traditional scheme acquires the ability to generate Chinese ink-style images after approximately 20 global rounds while our scheme requires about 40 rounds to attain the same proficiency. 

However, our scheme achieves a remarkable advantage in reducing the system cost in terms of the communication cost and time consumption for the model convergence, as shown in Fig. \ref{fig:case_study}(b). 
%provides the required system cost with respect to different training losses. 
To achieve an identical training loss of 0.02, the traditional and our schemes consume about 189 gigabytes and 1.3 gigabytes of data traffic, respectively. To achieve a training loss of 0.08, the traditional and our schemes take about 131 minutes and 2.9 minutes, respectively. Compared with the traditional scheme, our scheme can reduce both the communication costs and training latency by up to two orders of magnitude.  Fig. \ref{fig:case_study}(c) shows the image samples generated by different fine-tuned models on the client side. The results show that our scheme successfully enables a client to obtain the ability to generate Chinese ink-style paintings.

%\section{NUMERICAL RESULTS AND DISCUSSION}
%This section analyzes the performance of the proposed federated learning-aided AIGC approach through numerical simulations, comparing it with existing methods, e.g., centralized one.

\section{Future Research Directions}
To achieve the benefits of FL-empowered AIGC, there still exist several open and challenging issues.

\subsection{Incentive Mechanism Design for Federated Learning}
%To stimulate the contributions of data and resources used in the local model training, incentive mechanisms are necessitated to provide effective incentives for clients, which are willing to join the federated-learning-empowered AIGC. 

%In the federated pre-training and federated fine-tuning for AIGC, a simplified incentive mechanisms could be designed by a decision-maker to effectively motivate the clients and enable all participants to be considerately rewarded according to the quality of their real contributions. In this regard, mathematical tools of economic theory such as game theory, auction theory, contract theory can be adopted. However, aiming to improve the overall performance of federated learning, incentive mechanism design is not independent but should coordinate with other federated learning designs to achieve a centralized objective. For example, the incentive mechanism design for the clients and split point selection of an AIGC model are jointly optimized to enhance the federated learning in a split way. Facing the federated learning in a sequential way, the decision-maker considers a joint optimization scheme to design proper incentives while assigning proper processing orders to the clients. 

Incentive mechanisms are necessitated to provide effective incentives for clients, which are willing to join the federated-learning-empowered AIGC. Mathematical tools of economic theory such as game theory, auction theory, and contract theory can be adopted. %In this regard, game, auction and contract theories can be adopted.
Furthermore, the incentive mechanism design should coordinate with other FL designs to achieve a centralized objective. For example, the incentive mechanism design for the clients and split point selection of an AIGC model are jointly optimized to enhance the FL in a split approach. Facing the FL in a sequential approach, the decision-maker considers a joint optimization scheme to design proper incentives to the clients while assigning their processing orders. 

\subsection{Blockchain Assisted Decentralized AIGC Services}
As a decentralized ledger, %with advantages of traceability, immutability, transparency, and auditability,  
blockchain has been envisioned as the underlying technology to provide data security and transparent management for FL-aided AIGC.

On one hand, %blockchain is significant to improve the accountability, reliability, and incentive fairness  for  federated learning. 
blockchain supports decentralized and secure model storage for reliable FL by maintaining immutable records of all submitted local model updates and making them tamper-proof. %Blockchain also achieves a transparent approach for trustworthiness assessment and incentive management among the participating clients. 
Blockchain-based incentive mechanisms have been designed to  stimulate honest clients to contribute data and join the FL procedure. %where the contributions and incentives of all clients are recorded in a non-repudiation and tamper-resistance manner.
On the other hand, blockchain provides a valuable solution to protection of sensitive data of users and  effective management of AIGC products. %When accessing to AIGC services such as ChatGPT, the users may unintentionally disclose sensitive information to meet their needs. 
Blockchain can prevent the leakage of the information such as chat records of ChatGPT.  Blockchain also establishes a decentralized platform for the users to freely distribute,  authentic the ownership, and trade their content in the totally trustless environment. 

\subsection{Green AIGC Enabled By Semantic Communication}
%A typical application of deploying AIGC services in wireless networks is to generate qualified and customized digital content that is utilized to render graphics for Metaverse. This means that a Metaverse service provider can first collect sensing data of a user regarding the surrounding environment, and further perform the AIGC inference process to generate proper digital content according to the collected data, to render the 3D virtual world for the user. However,  a vast amount of the collected data generates for each user per second and the Metaverse service provider is difficult to receive and retrieve all raw data from multiple users to conduct the real-time rendering tasks. To cope with the dilemma,  semantic communication is leveraged as a promising technology to  extract available semantic information from the raw data and only transmit the semantic information with a small data size. This greatly reduces the data transmission time, saves the communication overhead and energy consumption in the system, which is important for the achievement of green AIGC service provision in wireless networks.  As a result, deploying AIGC services over semantic communication will be popular and meaningful to increase the user capacity and facilitate the content generation in a cost-effective manner. 
AIGC has been applied to generate digital content  to render graphics for Metaverse. A Metaverse service provider collects sensing data of a user and performs an AIGC inference based rendering task for the user. Semantic communication is further exploited to  extract available semantic information from the raw data and only transmit the semantic information with a small data size,  thereby saving the data transmission time and communication overheads in the system.
%process to generate proper digital content according to the collected data, and further renders the 3D virtual world for the user. %However,  a vast amount of the sensing data generates for each user per second and the Metaverse service provider is difficult to receive and retrieve all raw data to conduct the rendering tasks. To avoid transmitting a vast amount of the sensing data generated by each user, semantic communication is leveraged as a promising technology to  extract available semantic information from the raw data and only transmit the semantic information with a small data size,  thereby saving the data transmission time, communication overhead and energy consumption in the system. inherently %A semantic communication system trains a semantic encoder/decoder model to  extract/recover the semantic information. The conventional end-to-end semantic encoder/decoder model training is centralized and inherently vulnerable to attacks such as reconstruction attack, where an adversary knows the parameters of the encoder/decoder model to recover the raw data. 
In particular, FL can be leveraged to decentralize the semantic encoder/decoder model training   while improving the accuracy of  the semantic information extraction by using more training data.

%
%A semantic communication system needs to train a semantic encoder/decoder model to  effectively extract/recover the semantic information at the encoder/decoder. The conventional end-to-end semantic encoder/decoder model training is centralized and inherently vulnerable to attacks such as reconstruction attack, where an adversary knows the parameters of the encoder/decoder model and tries to recover the raw data. In this regard, federated learning can be adopted to decentralize the encoder/decoder model training to the multi-user model training to alleviate the security
%vulnerability while improving the accuracy of the semantic information extraction by using more training data and keeping the training data on the local storage. Thus, the application of federated learning is simultaneously beneficial to both the AIGC and semantic communication when deploying the green AIGC services over semantic communication.

\subsection{Personalized AIGC Based on Edge Intelligence}
Edge intelligence relies on exploiting the AI based decision-making capability for promoting the convergence of caching, computing and communication at the network edge and serves different users in the AIGC inference services.

%supports massive AIGC inference services in wireless networks according to different preferences of different users.

Many research efforts have been conducted to apply the edge intelligence to tackle the tradeoff between the model performance and resource consumption when multiple clients join the FL procedure.   %Different schemes that study how to jointly optimize the client selection, channel allocation, computing resource allocation, power control and the number of local iterations, are proposed to optimize the federated learning over wireless networks according to different purposes. and AIGC inference services are simultaneously required,
When a part of AIGC models are cached, the multi-user inference service delivery aims to  improve the overall user satisfaction while reducing the inference latency and resource consumption. The existing optimization schemes for FL can be modified to take into account the personalized requirements for the quality of the AIGC inference services.  However, it is rather difficult to quantitatively evaluate the impacts of the above optimization decisions on the performance metrics of the AIGC inference services, e.g.,  quality of the generated content, latency and reliability of content delivery and model hit ratio. %Therefore, learning based solution methodologies show great potential in addressing the tradeoff between inference accuracy, latency, and resource consumption in the incomplete information scenario.
\section{Conclusion}
The centralized training in the AIGC processes raises a challenging issue  for deploying AIGC services in wireless networks. To address this issue, we proposed FL to decentralize the model training procedures regarding AIGC, which can improve learning efficiency and achieve privacy protection for AIGC. We analyzed the issues remained on the integration of FL and AIGC, and then proposed instrumental FL-based techniques to implement the FL procedure including the  parallel, split and sequential approaches. As a case study, we  adopted FL in a sequential approach to perform the federated fine-tuning for a stable diffusion model. Numerical results verified the superiority of the proposed scheme.  Finally, we outlined the open research issues of FL-empowered AIGC.

\bibliography{myreference}

\end{document}